# Octahedral tilting and ferroelectricity in $RbANb_2O_7$ (A = Bi, Nd) from first principles

Hyunsu Sim and Bog G. Kim[*]

*Department of Physics, Pusan National University, Pusan, 609-735, South Korea*

The effects of octahedral tilting of $RbANb_2O_7$ (A = Bi, Nd) compounds was studied using density-functional theory. In this compound, the structural phase transition was correlated with two octahedral tilting modes ($a^-a^-c^0$ tilting and $a^0a^0c^+$ tilting), and magnitude of the octahedral tilting mode was analyzed in the optimized structure. The theoretical results correlated well with the recent experimental results on the ferroelectricity of $RbBiNb_2O_7$. The hybrid improper ferroelectricity resulting from the coupling of two octahedral tilting modes and off center displacement mode was analyzed by group theory and symmetry mode analysis. The detailed relationship of the tilting modes to the structural phase transition and the detailed physical properties of ferroelectricity are also presented.

*Electronic Address: boggikim@pusan.ac.kr

Recently, there has been considerable interest in lead-free ferroelectric materials, not only for technological importance, such as environmental issues, but also for their scientific importance [1-4]. Many ferroelectric oxides form $ABO_3$ perovskites with a 3D network of corner sharing octahedra [5]. The connectivity of octahedra can be modified to form a layered structure [6-13]. The Ruddlesden-Popper series, $A_2[A'_{n-1}B_nO_{3n-1}]$ is one of the most well-known examples [6, 7]. Another two series of layered perovskite is the Dion-Jacobson series [8, 9] and Aurivillius series [10, 11] with the chemical formulae, $A[A'_{n-1}B_nO_{3n+1}]$ and $Bi_2O_2[A'_{n-1}B_nO_{3n+1}]$, respectively. In a layered perovskite oxide, many lead-free ferroelectric candidates have been proposed [2, 12, 13].

$AA'B_2O_7$, which exhibits high temperature ferroelectricity without Pb, is a potential candidate for next the generation ferroelectric materials with a modified perovskite structure of layered octahedra [14 - 16]. This formula belongs to the structural class of the Dion-Jacobson series [8, 9]. In these materials, the B ion forms an octahedron with neighboring oxygen atoms and the $BO_6$ octahedra are linked to form a double layer. Each double layer is separated by an A atom and the A' atom imparts structural rigidity to the $BO_6$ octahedra.

As reported in early experimental and theoretical work on the Dion-Jacobson $CsBiNb_2O_7$ compound, one of the most important ingredients is octahedral tilting [17-20]. The room temperature structure of $CsBiNb_2O_7$ is a P$mc2_1$ orthorhombic structure [21] with $a^-a^-c^+$ tilting of the Glazer notation [21-24]. C. J. Fennie and Karin M. Rabe predicted the ferroelectric phase of $CsBiNb_2O_7$ using first-principles calculations [18]. Their calculation clearly showed that the coupling of the zone center ferroelectric mode and zone boundary octahedral tilting mode is the key origin for ferroelectric realization of a $CsBiNb_2O_7$ material. Experimentally, the ferroelectric nature of $CsBiNb_2O_7$ is unclear because of the large ionic conductivity of this material [19].

Octahedral tilting also plays important roles for various perovskite derived systems [25-28]. Bousquet *et al.* [25] reported that by layering perovskites in an artificial superlattice $SrTiO_3/PbTiO_3$, polarization can arise from the coupling of two octahedral tilting modes. N. A. Benedek and C. J. Fennie [26] also introduced the hybrid improper ferroelectricity concept in $(ABO_3)_2(AO)$ layered perovskites and reported that polarization, P, arises from octahedral tilting. Many studies have been devoted to octahedral rotation-induced ferroelectricity in cation ordered perovskites, $(ABO_3)/(A'B'O_3)$ systems [27-29]. In this sense, ferroelectricity in a Dion-Jacobson compound can be explained easily by hybrid improper ferroelectricity. More recently, B. W. Li et al, reported the experimental realization of high Tc ferroelectric behavior of a $RbBiNb_2O_7$ bulk material [30]. The ferroelectric transition temperature of 945 °C was confirmed by temperature dependent dielectric constant experiment, Raman experiments, and piezoelectric constant experiments.

Here we reports the results of a detailed analysis of the octahedral tilting and ferroelectric phase transition of $RbANb_2O_7$ (A = Bi and Nd) materials using density functional theory. The structure was optimized using first-principles calculations. Quantitative analysis was applied to octahedral tilting.

Next, the phonon dispersion curve was also calculated using frozen phonon method. Each octahedral tilting mode was analyzed quantitatively using first-principles data and mode analysis with group theory. The electronic structure and ferroelectric properties of these compounds are also presented and explained.

First principles calculations with the generalized gradient approximation (GGA) [31] were performed using density functional theory and the projector-augmented-wave method, as implemented in VASP [32, 33]. The following valence electron configurations were considered: $4s^24p^65s^1$ for Rb, $5d^{10}6p^26p^3$ for Bi, $5s^25p^65d^16s^2$ for Nd, $4p^64d^45s^1$ for Nb, and $2s^22p^4$ for O. Electronic wave functions were expanded with plane waves up to a kinetic-energy cutoff of 400 eV except for structural optimization, where a kinetic energy cutoff of 520 eV has been applied to reduce the effects of the Pulray stress. Momentum space integration was performed using $6 \times 6 \times 3$ Gamma-centered Monkhorst-Pack k-point mesh [34]. With the given symmetry of the perovskite imposed, the lattice constants and internal coordinates were fully optimized until the residual Hellmann-Feyman forces became smaller than $10^{-5}$ eV/Å. The total energy was calculated as a function of the fixed volume to obtain the equation of states. At each given volume, the internal atomic coordinate was fully optimized. To obtain phonon dispersion curve and phonon partial density of state, frozen phonon calculation was applied to a $2 \times 2 \times 2$ supercell (containing 176 atoms in a superlattice configuration) using the PHONOPY program [35]. Spontaneous polarization was obtained using the Berry phase technique [36]. The ISOTROPY and AMPLIMODES programs were utilized to check the group subgroup relationship and quantify octahedral tilting [37, 38].

Fig. 1 shows the optimized orthorhombic structure (SG #26 P$mc2_1$) of RbANb$_2$O$_7$ (A = Bi, Nd). The large pink sphere represents a Rb ion and the small yellow sphere represent Bi or Nd ion. Two NbO$_6$ octahedral are linked in the c-axis and they have significant distortion and tilting. Fig. 1(b) and (c) show the total energy of the tetragonal (without any tilting, P$4/mmm$) and orthorhombic phases for the Bi and Nd case, respectively. For a given volume, all atomic coordinates and cell shapes were relaxed to obtain each data point. Orthorhombic distortion reduces the total energy of the system and the equilibrium volume of the orthorhombic phase is smaller than the tetragonal phase. Tables 1 and 2 summarize the optimized cell parameters as well as atomic coordinates, respectively. The energy difference between the two phases was 1.15 and 0.64 eV/unit cell for the Bi and Nd compounds, respectively.

In order to see the phase stability of the tetragonal phase, the phonon dispersions were calculated for the tetragonal phase using the finite difference method with supercell approach. Fig. 2(a) and (b) show the phonon dispersion curve and the phonon partial density of state (PDOS) for the Bi and Nd compounds. In both compounds, phonon dispersion has a significant imaginary part throughout the Brillouin zone, which means that the tetragonal phase is unstable against a symmetry lowering phase transition to the orthorhombic phase. On the other hand, the detailed properties were significantly different for both compounds. First of all, the R-S phonon dispersion line appears to be similar for the two compounds but Gamma point instability is quite large for the Bi compound compared to the Nd compound (Fig. 2(a) and (b)). From a PDOS plot, the instability of Bi ion (Orange color in Fig. 2(a)) was larger than that of the Nd ion (Red color in Fig 2(b)). Note that the main contribution of the imaginary part of the phonon comes from oxygen ions (Black line in Fig. 2(a) and (b)), which is related to oxygen octahedral tilting. The Nb ion inside the octahedra makes a significant contribution to the imaginary part of the phonon (Green line in Fig. 2(a) and 2(b)), but the Rb ion contribution to the imaginary part of phonon is quite small compared to that of the other ions (Blue line in Fig. 2(a) and 2(b)).

Phonon instability in the tetragonal phase can be analyzed quantitatively by group theory using the ISOTROPY and AMPLIMODES programs [37, 38]. As an input for the reference structure in AMPLIMODES, we adapted a fully relaxed structure for the tetragonal and orthorhombic phase. Figure 3(a) shows the [001] projection of RbANb$_2$O$_7$ without any tilting and distortion. In the tetragonal phase, three phonon modes are responsible for the phonon instability. The $\Gamma_{5-}$ mode is the zone center mode associated with the displacement of Rb and Bi ions (shown in Fig. 3(b)) and M$_{3+}$ (M$_{2+}$ notation in Nd compound) and M$_{5-}$ are two zone boundary modes related to the tilting of $a^0a^0c^+$ and $a^-a^-c^0$ Glazer notation. Their [001] projections are depicted in Fig. 3(b), (c), and (d). We have calculated the total energy difference as a function of three modes for Bi (Fig. 3(e)) and Nd (Fig. 3(f)) compounds. The total energy decreases with increasing amplitude of each mode for small amplitude, which is related to the phonon instability of each mode. Finally, with large amplitude, the total energy increases again to show a double-well potential energy landscape. The minimum energy points for each mode are the stable energy, and the stable orthorhombic phase can be obtained by combining each amplitude. The total energy landscape in Fig. 3(e) and (f) is closely related to the phonon instability diagram shown in Fig. 2. For the $\Gamma_{5-}$ mode, the Bi compound showed a large energy change, whereas the Nd compound showed a small energy change, which is related to the magnitude of the zone center mode instability. The largest amplitude mode for Bi and Nd was M$_{3+}$ and M$_{5-}$ mode, respectively, whereas in terms energy stability, the $\Gamma_{5-}$ and M$_{5-}$ modes are most significant for Bi and Nd compounds, respectively. Tables I and II provide the detailed information of structural parameters and tilting mode analysis.

Having the detailed structural information of the orthorhombic phase and the corresponding tilting analysis, we are now in a position to correlate electronic properties of RbANb$_2$O$_7$ (A = Bi, Nd). The ferroelectric properties of the two materials were studied using the Berry Phase method [36]. Fig. 4 (a) presents the results of an analysis of the polarization of RbNdNb$_2$O$_7$. The spontaneous polarization of a fully distorted orthorhombic phase by the electronic calculation is the multiple values of the true spontaneous value plus an integer multiple of the polarization quanta, as explained by N. Spaldin [39]. The spontaneous polarization can be obtained from the plot in Fig. 4(a). The spontaneous polarization of the Bi and Nd compound was 35.72 μC/cm$^2$ and 24.53 μC/cm$^2$, respectively. Note that the experimental spontaneous polarization value of the Bi compound with a polycrystalline form was approximately 10 μC/cm$^2$ [30]. In terms of the tilting angle (shown in Table I), the c$^+$ tilting angle for the Bi compound was similar to that of the Nd compound (10.99 vs. 10.72), whereas the a$^-$ tilting angle for the Bi compound was smaller than that of the Nd compound (5.98 vs. 7.25). In addition, amplitude analysis of the tilting mode showed similar results, as shown in Table II (1.0671 vs. 1.0402 for a$^0$a$^0$c$^+$, and 0.8404 vs. 1.1604 for a$^-$a$^-$c$^0$). On the other hand, the structural transition from a tetragonal to orthorhombic phase is associated with three main modes with a large amplitude, a$^0$a$^0$c$^+$ tilting, a$^-$a$^-$c$^0$ tilting, and the opposite displacement of two A-site cations (RbBi and RbNd). In terms of the zone center ferroelectric instability ($\Gamma_{5-}$ mode), the RbBi compound is more efficient than the RbNd compound (0.7304 vs. 06025 in Table II). In other words, the total structural distortion for the RbNd compound is larger than that of the RbBi compound, the structural distortion related to the ferroelectric instability is larger for the RbBi compound.

Figure 5(a) and (b) depict the electronic structures (band structure and density of the states) of RbANb$_2$O$_7$ in the orthorhombic phase. The overall band structures of two compounds were similar in shape. The valence band was quite flat throughout the Brillouin zone, which means a large effective mass of the hole, and the valence band maximum (VBM) is located in (0.0, 1/3, 1/3) for the Bi compound and in the (0.0. 0.5, 0.0) Y point in the Nd compound. The conduction band was quite flat along the Z and Γ point (the minimum is located in the Γ point) but showed strong dispersion along other the Brillouin zone, suggesting an anisotropic effective mass of the conduction electrons near the conduction band minimum. The band gap (indirect) from our GGA of two compounds in the orthorhombic phase was 2.500 and 2.428 eV for Bi and Nd compounds, respectively. The atom projected density of the state showed a clear difference in the two compounds. The bottom of the conduction band was nearly determined by Nd atoms, hybridized with surrounding oxygen. The effect of the Rb atom in the conduction band and valence band was small, which can be understood from a structural point of view (Fig. 1(a)). The valence band of the Bi compound was determined mainly by Bi atoms, whereas that of the Nd compound was determined mainly by Nd and O atoms.

Figure 6(a) and (b) show the valence band charge density in the Bi-O (Nd-O) plane of the orthorhombic phase of the two compounds. The charge density of the two bands consisting of the valence band maximum was plotted for clarity. The contribution of Bi atoms to the valence band was quite strong and has an asymmetric shape. The reason for the asymmetry is strongly related to the lone pair s orbital of Bi atoms. Because of such asymmetry, the oxygen band of the Bi compound was quite different in shape compared to the Nd compound. The conduction band charge density in the Nb-O plane was plotted for two bands consisting of the conduction band maximum (Fig. 6(c) and (d)). The cross shape in the charge density plot was attributed to the d orbital of Nb atoms and the dumbbell shape is due to the oxygen p orbital. The Nd orbital was similar in the two compounds, whereas there was a small difference in the oxygen orbital. The difference in the two compounds was quite small, suggesting that the oxygen octahedra are quite insensitive to the structure.

In summary, we have examined the structural and electronic properties of ferroelectric compound $RbANb_2O_7$ (A = Bi, and Nd) using first-principles calculation. Detailed octahedral tilting analysis of two compounds in orthorhombic phase (SG #26 P*mc2$_1$*) was presented. The octahedral tilting modes are closely related to structural stability in the orthorhombic phase, and such tilting and A-site ordering of the Rb and A (Bi and Nd) atoms are important for the ferroelectric properties. The magnitude of the octahedral tilting of the Nd compound was larger than that of the Bi compound and the overall distortion from the tetragonal phase was also large for the Nd compound. However, the ferroelectric polarization was larger for the Bi compound, which is closely related to the magnitude of the $\Gamma_{5-}$ zone center symmetry mode. Partial charge density analysis also revealed the important role of Bi atoms in the ferroelectric phase. Coupling between the tilting modes and ferroelectric mode in the $RbANb_2O_7$ (A = Bi, and Nd) opens new engineering possibilities for $NbO_6$ octahedra-containing layered perovskites.

This study was supported by NSF of Korea (NRF-2013R1A1A2004496). Computational resources have been provided by KISTI Supercomputing Center (Project No. KSC-2013-C1-029).

**Figure Captions**

Figure 1. (a) P$mc2_1$ orthorhombic structure with $a^-a^-c^+$ octahedral tilting of RbANb$_2$O$_7$ (A = Bi, Nd). Total energy vs. volume of tetragonal and orthorhombic phase of (b) RbBiNb$_2$O$_7$ and (c) RbNdNb$_2$O$_7$.

Figure 2. Phonon dispersion curve and phonon partial density of state of P4/*mmm* tetragonal phase of (b) RbBiNb$_2$O$_7$ and (d) RbNdNb$_2$O$_7$. The R point is (0.5, 0.5, 0.5) and S point is (0.5, 0.5, 0) in the Brilloiun zone.

Figure 3. [001] projection of RbANb$_2$O$_7$ (A = Bi, Nd) (a) without tilting or distortion (tetragonal phase), (b) with $\Gamma_{5-}$ mode, (c) with $M_{3+}$ (or $M_{2+}$) mode, and (d) with $M_{5-}$ mode. Note that the A-site off centering can be seen in $\Gamma_{5-}$ mode. The total energy difference as a function of each mode amplitude in (e) RbBiNb$_2$O$_7$ and (f) RbNdNb$_2$O$_7$.

Figure 4. (a) Berry phase polarization calculation from undistorted tetragonal phase to a fully distorted orthorhombic phase of RbBiNb$_2$O$_7$. P$_Q$ is the polarization quantum and Ps is the spontaneous polarization of the P$mc2_1$ orthorhombic structure. (b) Spontaneous polarization in RbANb$_2$O$_7$ (A = Bi, Nd).

Figure 5. Band structure and partial density of state of the P$mc2_1$ orthorhombic phase of (a) RbBiNb$_2$O$_7$ and (b) RbNdNb$_2$O$_7$.

Figure 6. Valence band charge density in the Bi-O (Nd-O) plane of the P$mc2_1$ orthorhombic phase of (a) RbBiNb$_2$O$_7$ and (b) RbNdNb$_2$O$_7$. Conduction band charge density in the Nb-O plane of the P$mc2_1$ orthorhombic phase of (a) RbBiNb$_2$O$_7$ and (b) RbNdNb$_2$O$_7$. Each charge density scale was set from 0 to 70 % of the maximum charge density and the calculation was performed by two bands consisting of the valence band maximum or conduction band minimum.

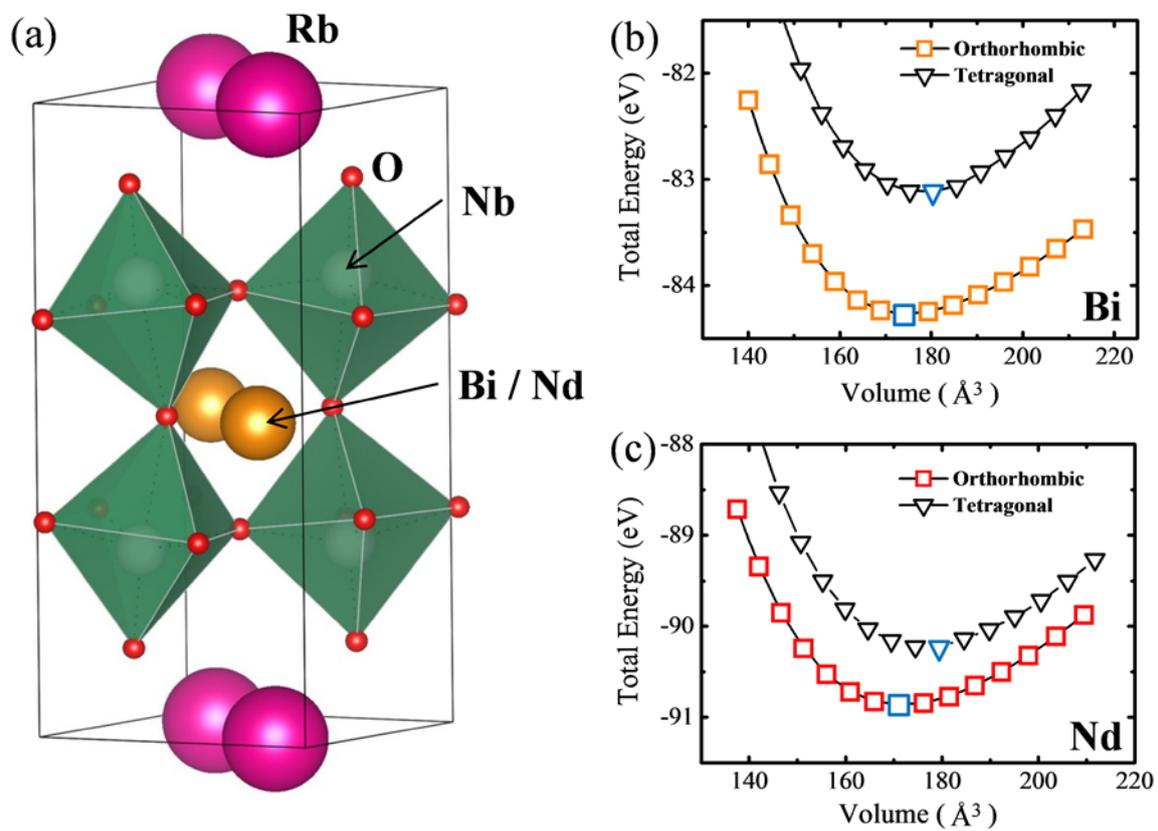

Figure 1 (Color Online) Sim and Kim

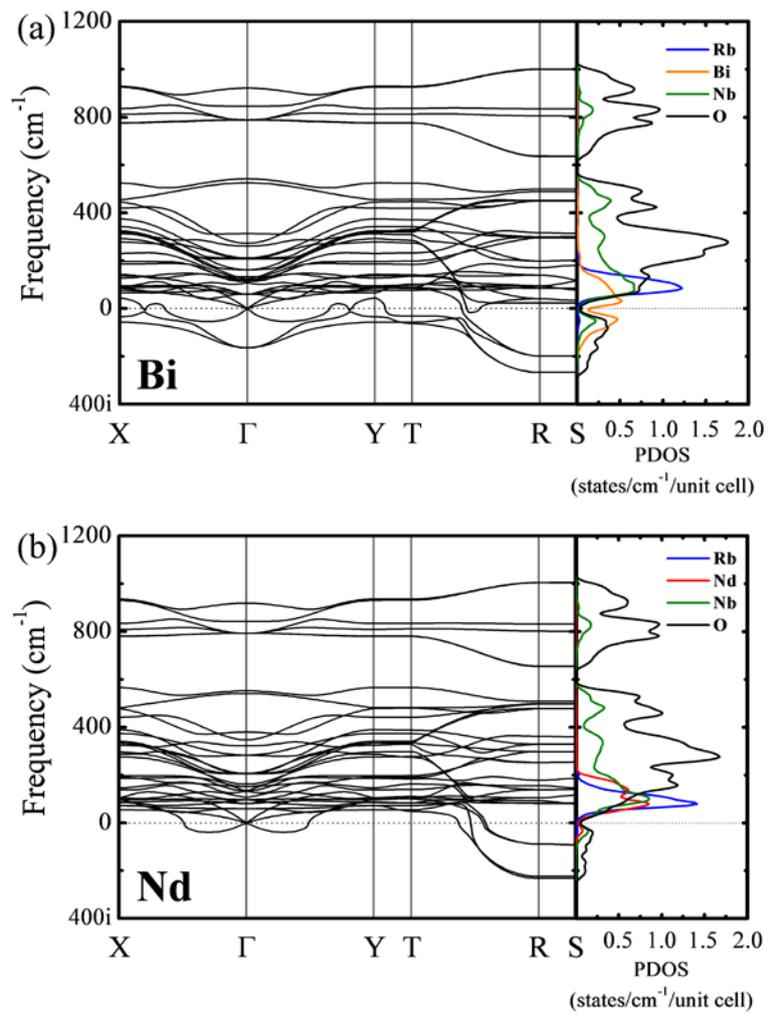

Figure 2 (Color Online) Sim and Kim

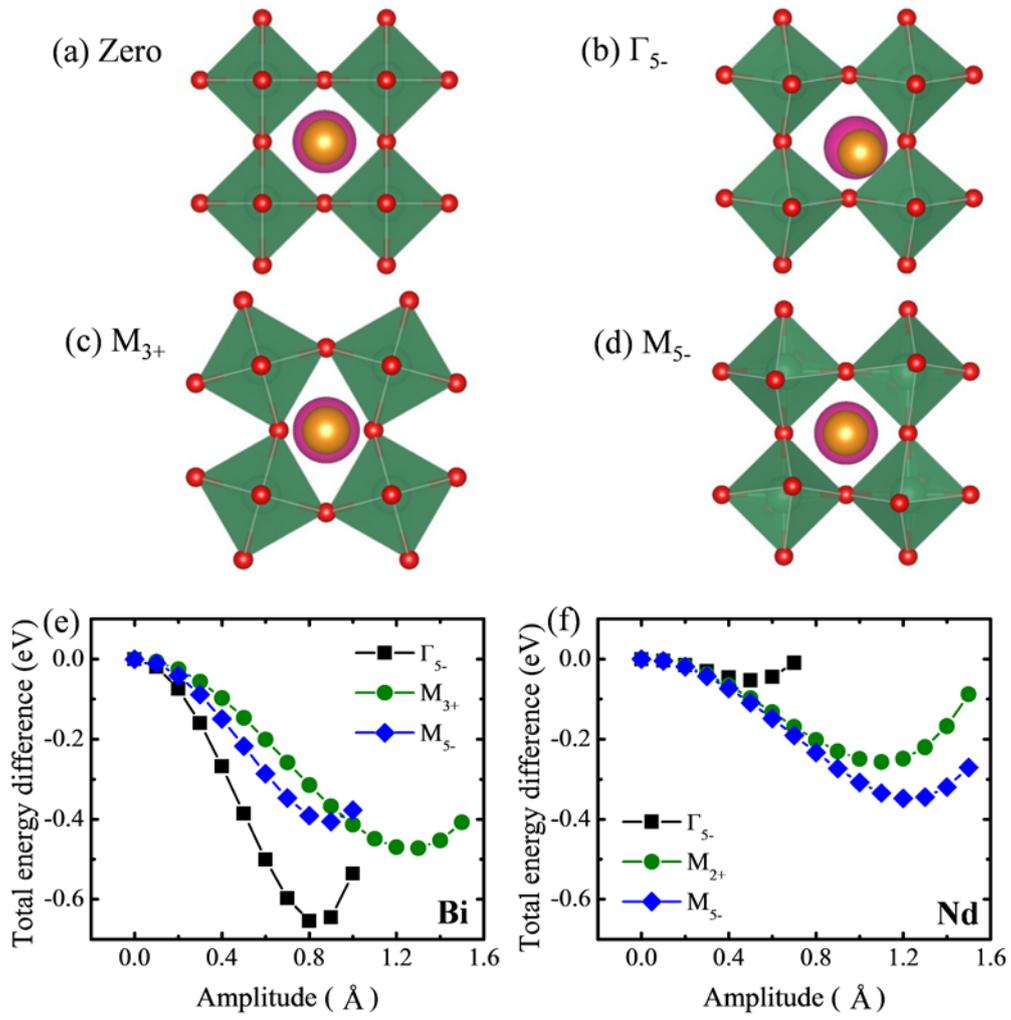

Figure 3 (Color Online) Sim and Kim

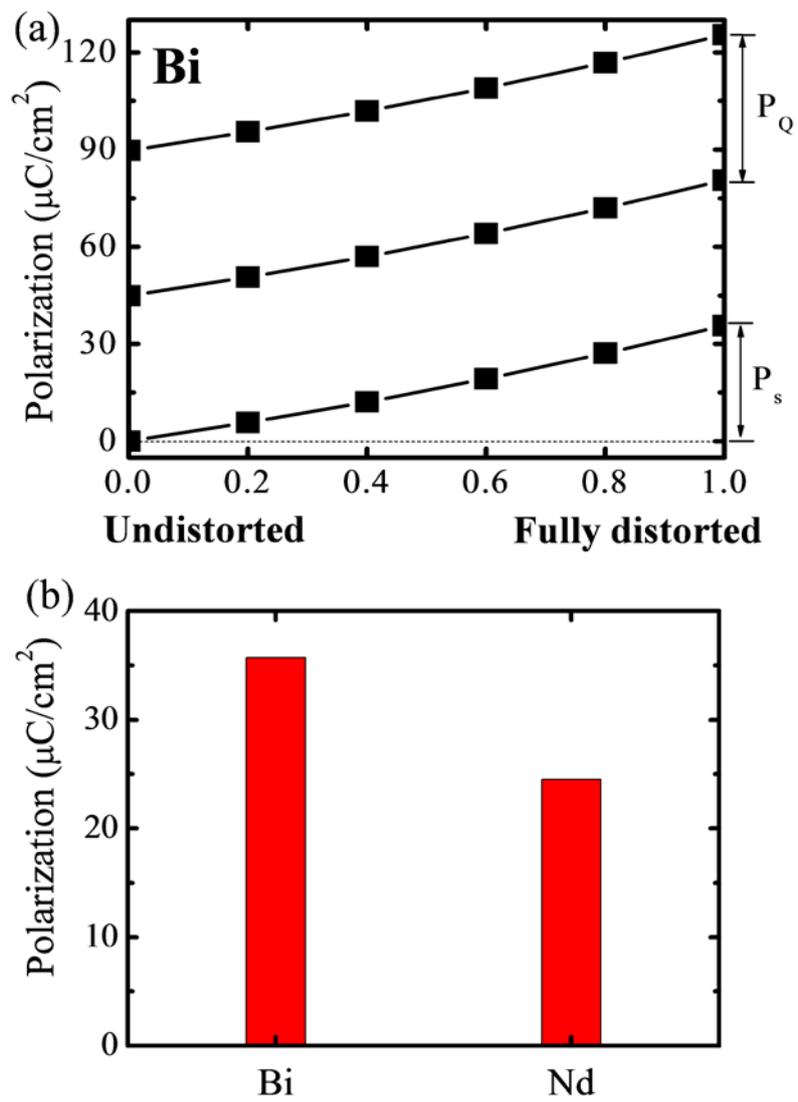

Figure 4 (Color Online) Sim and Kim

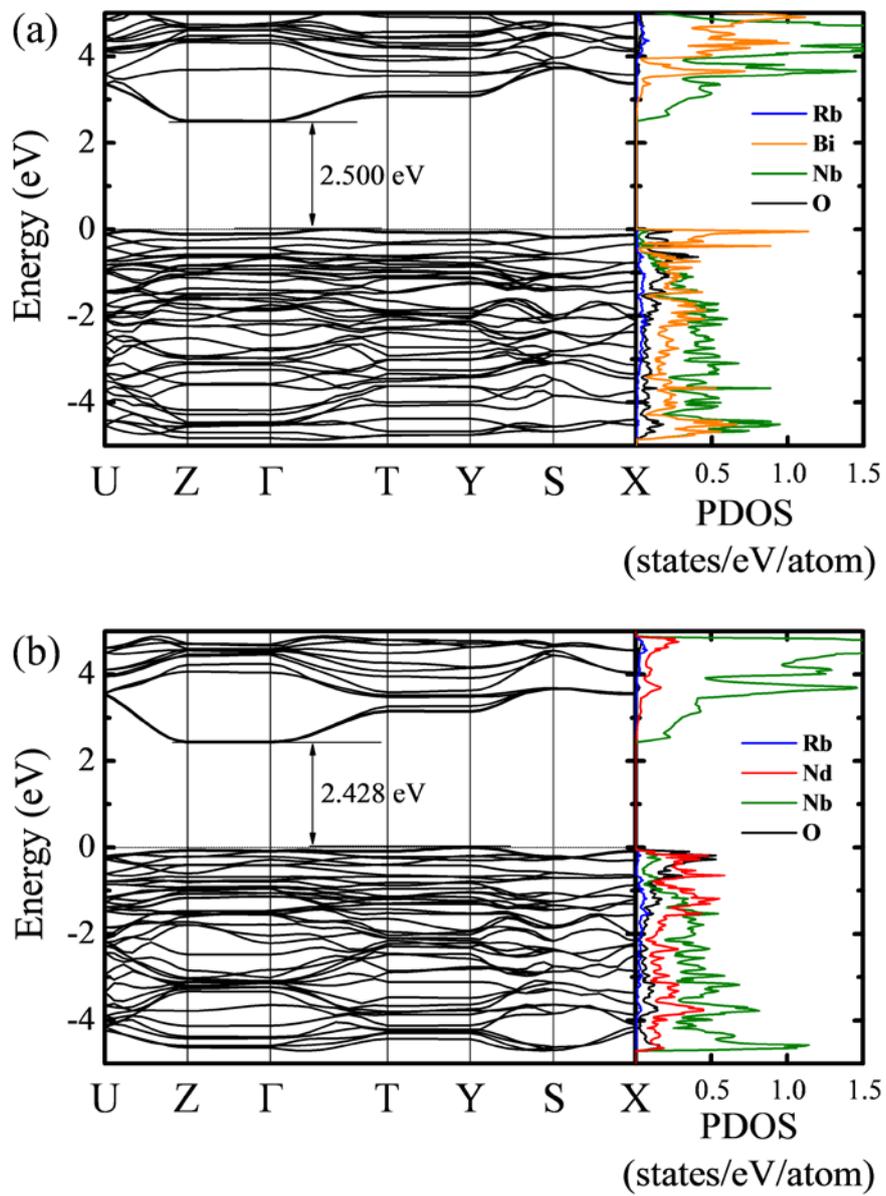

Figure 5 (Color Online) Sim and Kim

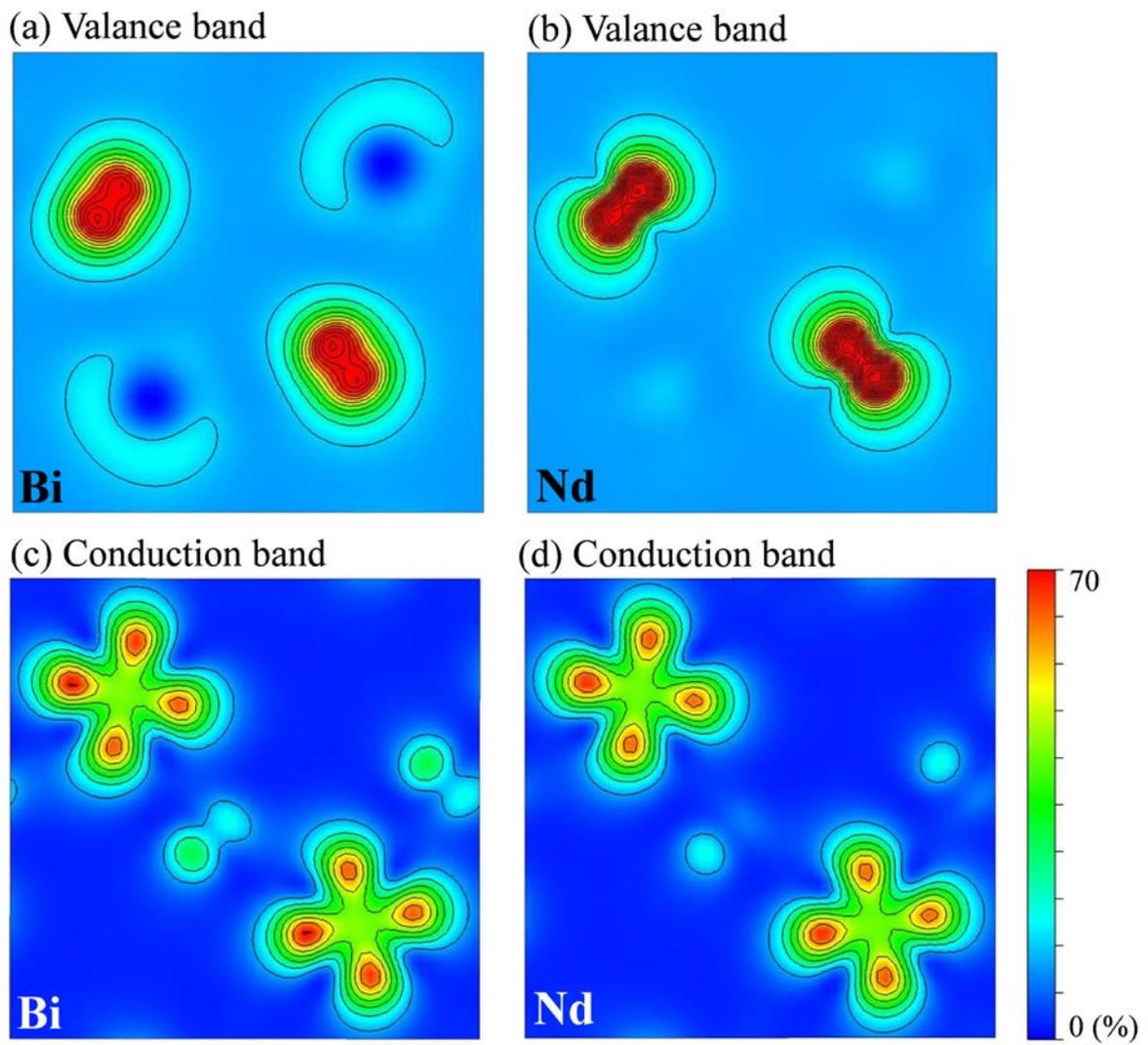

Figure 6 (Color Online) Sim and Kim

TABLE I. Structural and electronic properties of RbANb$_2$O$_7$ (A = Bi, Nd). The lattice constant, band gap energy, total energy, and polarization value are given for each space group (orthorhombic and tetragonal) of RbANb$_2$O$_7$ (A = Bi, Nd). Note that all orthorhombic phase have lower energy than the tetragonal phase and spontaneous polarization is observed in only the orthorhombic phase.

| Structure | Space group | Lattice parameter (Å) | | | Total energy (eV) | Eg (eV) | Tilting angle (deg) | | Polarization (μC/cm$^2$) |
|---|---|---|---|---|---|---|---|---|---|
| | | a | b | c | | | $a^-a^-c^0$ | $a^0a^0c^+$ | |
| RbBiNb$_2$O$_7$ | P4/mmm | 3.531 | 3.531 | 12.146 | -83.12 | 1.145 | 0 | 0 | 0 |
| | Pmc2$_1$ | 5.449 | 5.539 | 11.535 | -84.27 | 2.500 | 5.98 | 10.99 | 35.72 |
| RbNdNb$_2$O$_7$ | P4/mmm | 3.546 | 3.546 | 11.980 | -90.23 | 1.339 | 0 | 0 | 0 |
| | Pmc2$_1$ | 5.461 | 5.533 | 11.316 | -90.87 | 2.428 | 7.35 | 10.72 | 24.53 |

TABLE II. AMPLIMODES result of RbANb$_2$O$_7$ (A = Bi, Nd). K-vector, character, and amplitude of each mode are given in the orthorhombic phases. $\Gamma_{5-}$ mode is the ferroelectric mode, M$_{5-}$, M$_{3+}$, and M$_{2+}$ modes are the octahedral tilting modes.

| k-vector | Character | Amplitude in unit of angstrom (mode) | |
|---|---|---|---|
| | | RbBiNb$_2$O$_7$ | RbNdNb$_2$O$_7$ |
| (0,0,0) | Ferroelectric | 0.7304 ($\Gamma_{5-}$) | 0.6025 ($\Gamma_{5-}$) |
| (1/2,1/2,0) | Tilting ($a^-a^-c^0$) | 0.8404 (M$_{5-}$) | 1.1604 (M$_{5-}$) |
| (1/2,1/2,0) | Rotation ($a^0a^0c^+$) | 1.0671 (M$_{3+}$) | 1.0402 (M$_{2+}$) |

# Supplementary information

Hyunsu Sim[1], and Bog G. Kim[1,*]

*Department of Physics, Pusan National University, Pusan, 609-735, South Korea*

Some of detailed calculation results are summarized here. Other results can be easily obtainable using the parameters presented here. Input and Output files for some calculations can be provided upon e-mail request (boggikim@pusan.ac.kr).

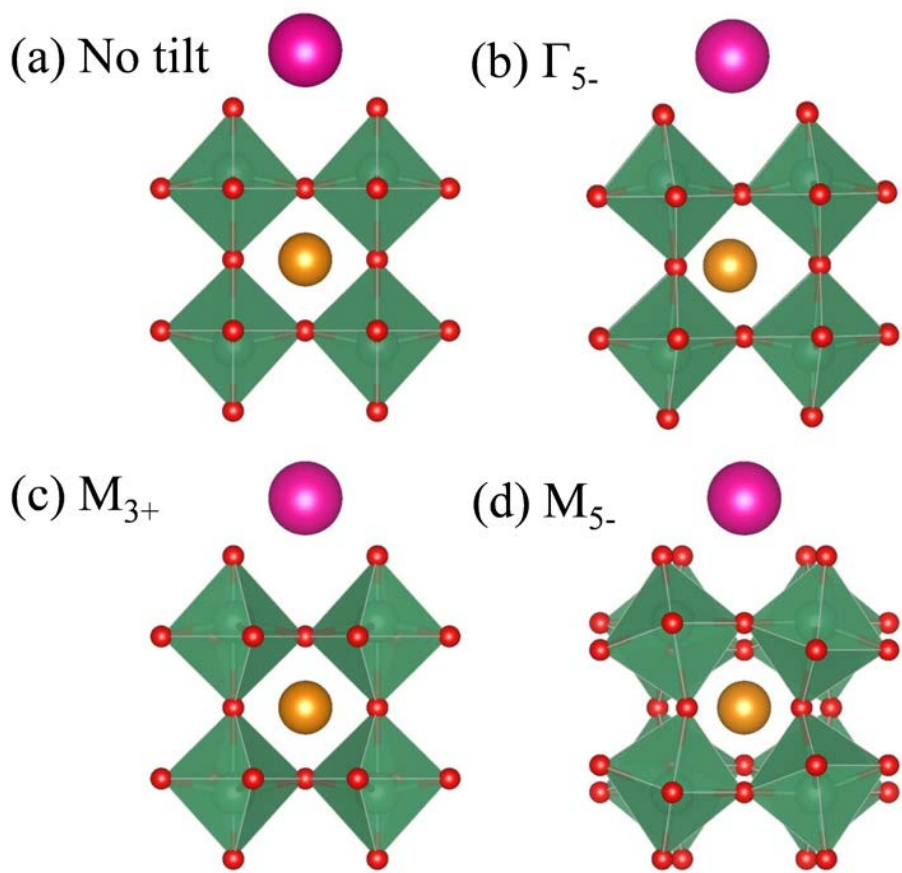

Figure S1. [100] projection of $RbANb_2O_7$ (A = Bi, Nd) (a) without tilting or distortion(tetragonal phase), (b) with $\Gamma_{5-}$ mode, (c) with $M_{3+}$ (or $M_{2+}$) mode, and (d) with $M_{5-}$ mode. Note that the A-site off centering can be seen in $\Gamma_{5-}$ mode.

TABLE S1. Wyckoff positions of atoms in tetragonal and orthorhombic phase of $RbANb_2O_7$ (A = Bi, Nd)

| Structure | Space group | SITE | Atom | x | y | z |
|---|---|---|---|---|---|---|
| RbBiNb$_2$O$_7$ | P4/mmm | 1d | Rb | 0.5000 | 0.5000 | 0.5000 |
| | | 1c | Bi | 0.5000 | 0.5000 | 0.0000 |
| | | 2g | Nb | 0.0000 | 0.0000 | 0.2307 |
| | | 1a | O | 0.0000 | 0.0000 | 0.0000 |
| | | 2g | O | 0.0000 | 0.0000 | 0.3799 |
| | | 4i | O | 0.0000 | 0.5000 | 0.1694 |
| | Pmc2$_1$ | 2b | Rb | 0.2497 | 0.7447 | 0.5000 |
| | | 2a | Bi | 0.2731 | 0.6909 | 0.0000 |
| | | 4c | Nb | 0.7535 | 0.7554 | 0.2106 |
| | | 2a | O | 0.6728 | 0.7944 | 0.0000 |
| | | 4c | O | 0.7797 | 0.7493 | 0.3658 |
| | | 4c | O | 0.0378 | 0.9826 | 0.1598 |
| | | 4c | O | 0.4401 | 0.5792 | 0.1882 |
| RbNdNb$_2$O$_7$ | P4/mmm | 1d | Rb | 0.5000 | 0.5000 | 0.5000 |
| | | 1c | Bi | 0.5000 | 0.5000 | 0.0000 |
| | | 2g | Nb | 0.0000 | 0.0000 | 0.2261 |
| | | 1a | O | 0.0000 | 0.0000 | 0.0000 |
| | | 2g | O | 0.0000 | 0.0000 | 0.3774 |
| | | 4i | O | 0.0000 | 0.5000 | 0.1654 |
| | Pmc2$_1$ | 2b | Rb | 0.2501 | 0.7371 | 0.5000 |
| | | 2a | Nd | 0.2489 | 0.7141 | 0.0000 |
| | | 4c | Nb | 0.7486 | 0.7558 | 0.2050 |
| | | 2a | O | 0.6710 | 0.7890 | 0.0000 |
| | | 4c | O | 0.8030 | 0.7462 | 0.3616 |
| | | 4c | O | 0.0373 | 0.9835 | 0.1449 |
| | | 4c | O | 0.4404 | 0.5759 | 0.1949 |

TABLE SII. Atomic displacement of normal modes (in relative units) with given symmetry in RbANb$_2$O$_7$ (A = Bi, Nd)

| RbBiNb$_2$O$_7$ | | Reference structure | | | $\Gamma_{5-}$ Ferroelectric | | |
|---|---|---|---|---|---|---|---|
| Lattice constant | | 5.4496 | 5.5388 | 11.535 | | | |
| Atom | wyckoff | x | Y | Z | dx | dy | dz |
| Rb | 2b | 0.2500 | 0.7500 | 0.5000 | 0.0000 | -0.0213 | 0.0000 |
| Bi | 2a | 0.2500 | 0.7500 | 0.0000 | 0.0000 | -0.0950 | 0.0000 |
| Nb | 4c | 0.7500 | 0.7500 | 0.2106 | 0.0000 | -0.0067 | 0.0000 |
| O | 2a | 0.7500 | 0.7500 | 0.0000 | 0.0000 | 0.0468 | 0.0000 |
| O | 4c | 0.7500 | 0.7500 | 0.3658 | 0.0000 | -0.0151 | 0.0000 |
| O | 4c | 0.0000 | 0.0000 | 0.1739 | -0.0151 | 0.0283 | 0.0000 |
| O | 4c | 0.5000 | 0.5000 | 0.1739 | -0.0151 | 0.0283 | 0.0000 |
| RbBiNb$_2$O$_7$ | | $M_{3+}$ Rotation ($a^0a^0c^+$) | | | $M_{5-}$ Tilting ($a^-a^-c^0$) | | |
| Atom | wyckoff | dx | dy | dz | dx | dy | dz |
| Rb | 2b | 0.0000 | 0.0000 | 0.0000 | -0.0004 | 0.0000 | 0.0000 |
| Bi | 2a | 0.0000 | 0.0000 | 0.0000 | 0.0275 | 0.0000 | 0.0000 |
| Nb | 4c | 0.0000 | 0.0000 | 0.0000 | 0.0041 | 0.0000 | 0.0000 |
| O | 2a | 0.0000 | 0.0000 | 0.0000 | -0.0918 | 0.0000 | 0.0000 |
| O | 4c | 0.0000 | 0.0000 | 0.0000 | 0.0353 | 0.0000 | 0.0000 |
| O | 4c | 0.0455 | -0.0455 | 0.0000 | 0.0000 | 0.0000 | -0.0169 |
| O | 4c | -0.0455 | 0.0455 | 0.0000 | 0.0000 | 0.0000 | 0.0169 |

| RbNdNb$_2$O$_7$ | | Reference structure | | | | $\Gamma_{5-}$ Ferroelectric | |
|---|---|---|---|---|---|---|---|
| Lattice constant | | 5.4614 | 5.5333 | 11.316 | | | |
| Atom | wyckoff | x | y | z | dx | dy | dz |
| Rb | 2b | 0.7500 | 0.7500 | 0.5000 | 0.0000 | -0.0384 | 0.0000 |
| Nd | 2a | 0.7500 | 0.7500 | 0.0000 | 0.0000 | -0.0767 | 0.0000 |
| Nb | 4c | 0.2500 | 0.7500 | 0.2049 | 0.0000 | -0.0074 | 0.0000 |
| O | 2a | 0.2500 | 0.7500 | 0.0000 | 0.0000 | 0.0477 | 0.0000 |
| O | 4c | 0.2500 | 0.7500 | 0.3616 | 0.0000 | -0.0234 | 0.0000 |
| O | 4c | 0.0000 | 0.5000 | 0.1699 | -0.0185 | 0.0323 | 0.0000 |
| O | 4c | 0.5000 | 0.0000 | 0.1699 | -0.0185 | 0.0323 | 0.0000 |
| RbNdNb$_2$O$_7$ | | $M_{2+}$ Rotation ($a^0a^0c^+$) | | | $M_{5-}$ Tilting ($a^-a^-c^0$) | | |
| Atom | wyckoff | dx | dy | dz | dx | dy | dz |
| Rb | 2b | 0.0000 | 0.0000 | 0.0000 | -0.0001 | 0.0000 | 0.0000 |
| Nd | 2a | 0.0000 | 0.0000 | 0.0000 | 0.0010 | 0.0000 | 0.0000 |
| Nb | 4c | 0.0000 | 0.0000 | 0.0000 | 0.0012 | 0.0000 | 0.0000 |
| O | 2a | 0.0000 | 0.0000 | 0.0000 | 0.0681 | 0.0000 | 0.0000 |
| O | 4c | 0.0000 | 0.0000 | 0.0000 | -0.0457 | 0.0000 | 0.0000 |
| O | 4c | 0.0455 | -0.0455 | 0.0000 | 0.0000 | 0.0000 | -0.0215 |
| O | 4c | -0.0455 | 0.0455 | 0.0000 | 0.0000 | 0.0000 | 0.0215 |